\documentclass[times,12pt]{article}
\usepackage[margin=1in]{geometry}
\usepackage[hidelinks]{hyperref}
\usepackage{lineno}
\usepackage{amsmath, amssymb}
\usepackage{hyperref}
\usepackage{graphicx}
\usepackage[round, comma]{natbib}
\usepackage{psfrag}
\usepackage[lofdepth,lotdepth]{subfig}
\usepackage{mathtools}
\usepackage{multirow, authblk}
\usepackage{multicol}
\usepackage{bm}
\usepackage{threeparttable, tablefootnote}
\usepackage{xcolor}
\usepackage{soul}
\extrafloats{100}
\usepackage{subfig}
\usepackage{chngcntr}
\usepackage{setspace}
\hypersetup{
    colorlinks,
    linkcolor={red!50!black},
    citecolor={blue!50!black},
    urlcolor={blue!80!black}
}

\begin{document}

\noindent \textbf{The influence of upper boundary conditions on molecular kinetic atmospheric escape simulations} \\
\noindent Shane R. Carberry Mogan$^{a, b, *}$, Orenthal J. Tucker$^c$, Robert E. Johnson$^{a, d}$ \\
\noindent $^a$NYU, New York, USA; $^b$NYU Abu Dhabi, Abu Dhabi, UAE; $^c$NASA GSFC, Maryland, USA; $^d$UVa, Virginia, USA; *Corresponding author at: Center for Space Science, New York University Abu Dhabi, Abu Dhabi, UAE; \textit{E-mail address}: ShaneRCM@nyu.edu. \\
\noindent \textbf{Abstract}: Molecular kinetic simulations are typically used to accurately describe the tenuous regions of the upper atmospheres on planetary bodies. These simulations track the motion of particles representing real atmospheric atoms and/or molecules subject to collisions, the object's gravity, and external influences. Because particles can end up in very large ballistic orbits, upper boundary conditions (UBC) are typically used to limit the domain size thereby reducing the time for the atmosphere to reach steady-state. In the absence of a clear altitude at which all molecules are removed, such as a Hill sphere, an often used condition is to choose an altitude at which collisions become infrequent so that particles on escape trajectories are removed. The remainder are then either specularly reflected back into the simulation domain or their ballistic trajectories are calculated analytically or explicitly tracked so they eventually re-enter the domain. Here we examine the effect of the choice of the UBC on the escape rate and the structure of the atmosphere near the nominal exobase in the convenient and frequently used 1D spherically symmetric approximation. Using Callisto as the example body, we show that the commonly used specular reflection UBC can lead to significant uncertainties when simulating a species with a lifetime comparable to or longer than a dynamical time scale, such as an overestimation of escape rates and an inflated exosphere. Therefore, although specular reflection is convenient, the molecular lifetimes and body's dynamical time scales need to be considered even when implementing the convenient 1D spherically symmetric simulations in order to accurately estimate the escape rate and the density and temperature structure in the transition regime.

\section{Introduction} \label{sec:intro}

In order to accurately describe the tenuous region near an atmosphere's exobase, it has been shown that solutions to the Boltzmann equation or molecular kinetic simulations, such as the direct simulation Monte Carlo (DSMC) method \citep{bird1994}, are required. Because such simulations can be computationally expensive, assumptions and simplifications are implemented to reduce run-times. One such simplification is to assume the upper atmosphere is on average spherically symmetric so that the steady-state radial structure and flow are calculated in 1D. Of course, atmospheres are not symmetric nor are their flows steady, thus an accurate description requires simulations that are  multi-dimensional and include transient processes (e.g., \citealt{carberrymogan2021, leblanc2017, lee2015, walker2012}). Such simulations, however, require large computer clusters and elaborate computational schemes to efficiently handle large batches of data, and can take years to develop. In contrast the 1D simulations can often be run on a modern desktop computer to obtain useful approximations to the upper atmospheric structure and processes. Indeed, 1D DSMC models have been applied to a wide variety of planetary atmospheres including complex and transient processes: gravity waves in Mars' atmosphere \citep{leclercq2020}; coupling to fluid models of Pluto's lower atmosphere \citep{erwin2013, tucker2012}; cooling in multi-component atmospheres induced via escaping H$_2$ \citep{carberrymogan2020, tucker2013}; thermal and non-thermal processes in Europa's atmosphere \citep{shematovich2005}; the Moon's early volcanic atmosphere \citep{tucker2020}; the collapse and reformation of Io's atmosphere during and after eclipse \citep{moore2009}; volatile loss from Kuiper Belt Objects \citep{johnson2015}; and cometary coma \citep{combi1996}; to name several. Furthermore, 1D DSMC simulations can be used to evaluate standard analytic models: e.g., Jeans escape \citep{volkov2011a}; the exobase approximation \citep{tucker2016}; and energy-limited escape \citep{johnson2013}. Because of their usefulness, this Note focuses on the effect of typical boundary conditions on the upper atmospheric structure, escape rate, and the required simulation times in 1D.

To describe the transition with altitude from the collision dominated regime to the collisionless regime in a single-component atmosphere \cite{merryfield1994} used a particle-based discrete velocity model to numerically integrate the nonlinear Boltzmann equation and, more recently, \cite{volkov2011a}, hereafter referred to as V11, used 1D DSMC simulations. Such simulations show that as a result of adiabatic expansion and the concomitant escape and cooling, the quasi steady-state can differ significantly from that predicted by solutions to fluid equations or to typical approximate analytic models. Moreover, since molecules can escape from below the nominal exobase and collisions can contribute to the escape of molecules above the nominal exobase, the often used Jeans escape rate calculated at the nominal exobase or the lower boundary can be significantly in error relative to accurate molecular kinetic simulations. V11 showed the difference between escape rates calculated using an approximate analytic model and via 1D DSMC simulations depended on the atmospheric parameters at the lower boundary: the Jeans parameter (i.e., the ratio of gravitational binding energy to thermal energy) and Knudsen number (i.e., the ratio of the mean free path for collisions to an atmospheric length scale), both of which are discussed in more detail below. Using an upward flux boundary condition, the lower boundary in these simulations can be an altitude in the atmosphere at which conditions are known or can be the physical surface as in V11. Whereas collisions near the lower boundary can cause particles to rapidly return to it, collisions at high altitudes can restrict escape or give particles additional energy to escape. This combination can produce escape rates either smaller or larger than the Jeans escape rate. More importantly, this difference between real and Jeans escape rates can depend on the placement and conditions at the upper boundary as discussed below. For particles that reached the upper boundary of the domain with an energy lower than the escape energy, \cite{merryfield1994} analytically calculated a time delay for them to eventually re-enter the domain through the upper boundary on ballistic trajectories. On the other hand, V11 implemented specular reflection, in which particles on ballistic trajectories that eventually return to the domain are instead immediately reflected back so as to shorten the time to reach steady-state. Although these models use different conditions at the upper boundary, they both ignore collisions above the upper boundary and, therefore, the steady-state results can be in agreement. 

There are of course other alternatives to these upper boundary conditions. For example, instead of implementing an analytic approximation for the time at which particles following ballistic trajectories above the upper boundary eventually return to the domain, the motion of these ballistic particles can be explicitly tracked until they return to the domain. \cite{leclercq2020} implemented this approach because they simulated a transient phenomena, wave activity, for which specular reflection is inadequate since it assumes the atmosphere is in steady-state and, as a result, would wash out any transient effects. A simpler example is to remove all particles that reach the prescribed upper boundary. This is often implemented in multi-dimensional models (e.g., \citealt{marconi2007}), where increasing the upper boundary or tracking ballistic particles until they eventually return to the atmosphere is computationally expensive. Moreover, in 1D tracking ballistic particles above the upper boundary is relatively simple because all particles that return to the domain return to the same upper layer, regardless of their angular coordinates. Conversely, in a 2D or 3D domain, the upper layer is broken up into angular components and thus the angular coordinates of the particles matter, making tracking them more complex, especially when implementing parallel processing whereby the multi-dimensional spatial domain is further divided up among individual processors and exchanging particle data can become a tedious and even computationally expensive process. In addition, when simulating an atmosphere comprised of a heavy species where the prescribed upper boundary is at a relatively high altitude (e.g., \citealt{austin2000, walker2010}), in the rare case that a particle crosses this boundary, its return to the domain can have a negligible influence on the atmospheric flow. However, removing all particles that cross the upper boundary can also be physically relevant in some instances, such as when it is set to the body's Hill sphere radius so that all particles that cross it are no longer gravitationally bound to the body of interest \citep{hill1878}.

The scaling described above does not apply to species-dependent molecular lifetimes nor dynamical time scales of the body. Since the simulations in this study require information from a particular planetary body and a particular species, here we use the properties of Callisto, the outermost Galilean satellite of Jupiter, and H$_2$, a long-lived species common to many early and present atmospheres (e.g., \citealt{carberrymogan2020, carberrymogan2021}; \citealt{tucker2013, tucker2020}). Therefore, although the results presented below are specific, the influence of the upper boundary conditions must be considered in all such simulations.

\section{Numerical Method}

Herein we apply the direct simulation Monte Carlo (DSMC) method \citep{bird1994} to simulate thermal escape in a 1D spherically symmetric domain. V11 showed that when simulating single-component atmospheres with DSMC the escape rates and structure can be scaled with the Jeans parameter, $\lambda_\mathrm{J,0}$, and the Knudsen number, Kn$_0$, calculated using the source parameter, the lower boundary, $r_0$. The former represents the ratio of the gravitational binding energy of the planetary object ($\frac{GMm}{r_0}$) to the thermal energy ($kT$) and the latter is a dimensionless parameter used to characterize the degree of rarefaction in the atmosphere. Here $G$ is the gravitational constant, $M$ is the mass of the planetary object, $m$ is the mass of the species, $k$ is the Boltzmann constant, and $T$ is the temperature. As discussed in V11, the Knudsen number at a given radial position, $r$, can be defined by two ways depending on the atmospheric curvature: the ratio of the local mean free path between collisions, $\ell(r)$, to either the local atmospheric scale height, $H(r) = \frac{k T(r) r^2}{G M m}$, here written Kn$(r)$, or to the source parameter, $r_0$, written as Kn$_0(r)$. 

Because the results in 1D can be roughly scaled using $\lambda_\mathrm{J,0}$ and Kn$_0$, the spherical body in these simulations is that of Callisto with a mass and radius of $M_C = 1.08 \times 10^{23}$ kg and $r_C = 2410$ km, respectively, with the lower boundary the physical surface, $r_0 = r_C$. Radial cells are generated from $r_0$ up to a chosen upper boundary, $r_\mathrm{max}$, which varies over the several simulations presented. The simulated atmospheres are produced by a thermal flux across the lower boundary. We assume the initial state is a vacuum, although that is not necessary, so that simulation particles representing atmospheric molecules are injected from $r_0$ into initially empty cells assuming a Maxwellian flux \citep{brinkmann1970} based on the temperature at the lower boundary, $T_0$. In this study, the atmospheric molecules are H$_2$, and the number of molecules represented by each particle is a function of the density at the lower boundary, $n_0$, and $T_0$, both of which can be varied relative to the source parameters. Particle motion is tracked in 3D Cartesian coordinates using a 4th-order Runge Kutta integration and is influenced by gravity and binary collisions. Here we simulate inelastic collisions between particles using the variable hard sphere (VHS) model \citep{bird1994}, where the redistribution of energy among kinetic (translational) and internal (rotational) modes is calculated using the Larsen-Borgnakke (LB) model \citep{larsen1974}. Although the scaling using hard sphere (HS) collisions, as in V11, is exact, it is only approximate for the more accurate VHS and LB models used here.

As these simulations step forward in time, particles' new radial positions are calculated. Those returning to the lower boundary can be treated in a number of ways depending on the molecule type, the surface composition and temperature, etc. Because we are focused on the upper boundary, here we simply remove particles that return to the surface. In simulating a steady-state atmosphere, conditions at the upper boundary are varied as follows:  

\noindent (1) \textbf{Reflect} \boldsymbol{$v < v_\mathrm{esc} (r_\mathrm{max})$}: particles that reach $r_\mathrm{max}$ with speeds greater than the escape speed, $v_\mathrm{esc}$, calculated at $r_\mathrm{max}$, $v_\mathrm{esc} (r_\mathrm{max}) = \sqrt{\frac{2 G M_C}{r_\mathrm{max}}}$, with an upward trajectory are removed, and all other particles are specularly reflected back into the domain as in V11; 

\noindent (2) \textbf{Reflect} \boldsymbol{$v < v_\mathrm{HS} (r_\mathrm{max})$}: particles that reach $r_\mathrm{max}$ with speeds that would allow them to reach the Hill sphere radius, $r_\mathrm{HS}$, if their upward trajectories were uninhibited above $r_\mathrm{max}$, $v_\mathrm{HS} (r_\mathrm{max}) = \sqrt{2 G M_C \left(\frac{1}{r_\mathrm{max}} - \frac{1}{r_\mathrm{HS}} \right)}$,  are removed, and all other particles are specularly reflected back into the domain;

\noindent (3) \textbf{Ballistic} \boldsymbol{$r > r_\mathrm{max}$}: particles that reach $r_\mathrm{max}$ with speeds greater than $v_\mathrm{esc} (r_\mathrm{max})$ are removed, while those with lower speeds are tracked along ballistic trajectories above $r_\mathrm{max}$ until they return to the domain; and 

\noindent (4) \textbf{Free Escape}: all particles that reach $r_\mathrm{max}$ are removed. 

Here $r_\mathrm{HS}$ is that of Callisto: $r_\mathrm{HS} = \left( \frac{M_C}{3 M_J} \right)^{1/3} d_{JC} \sim 20.8 r_C$, where $M_J$ = 1.898$ \times 10^{27}$ kg is the mass of Jupiter and $d_{JC} \sim 26.3 r_J$ is the distance from Callisto to Jupiter in units of Jupiter radii, $r_J = 71,492$ km. 

The escape rate is calculated by averaging the rate at which particles leave the domain according to the upper boundary conditions described above over a time interval much shorter than the total simulation time. After an average is taken, the escape rate is then reset for the subsequent interval. For the time intervals of this study, we took averages every $10^5$ steps, which equate to $\sim$7.5$\times 10^4$--3$\times 10^5$ s in the simulations presented below depending on the length of the time-step ($\sim$0.75--3 s), and there was a total of $10^7$ steps in each simulation. Steady-state is determined when the differences between macroscopic averages, such as escape rates, number density, and temperature, are negligible. As will be shown below, when implementing upper boundary condition (3), due to the long times it can take for particles following ballistic trajectories above $r_\mathrm{max}$ to return to the domain steady-state is still not reached after $10^7$ steps. Conversely, for all other upper boundary conditions steady-state is attained relatively rapidly, within the first few averages taken; however, the simulations are still run over a longer time to reduce statistical noise in the results. Note in a dynamical simulation, the averaging interval must be much shorter than the timescale of interest; e.g., a planetary body's orbital period, $t_\mathrm{orb}$. In the results shown below, when relevant, we normalize the simulation time to Callisto's orbital period, $t_\mathrm{orb} \sim 1.44 \times 10^6$ s, or present the atmospheric structure after 1 $t_\mathrm{orb}$ in order to stress the importance of the time it can take a simulation to reach steady-state relative to a physically relevant timescale and how the upper boundary conditions can affect this result. 

\section{Results} 

Molecular kinetics in 1D single-component H$_2$ model atmospheres at Callisto were simulated using the DSMC method. The nominal exobase, $r_x$, which acts as a reference point, is calculated to occur at either Kn$(r_x)\sim 1$ or Kn$_0(r_x)\sim 1$. Since the Knudsen number is a function of the local density and temperature, and both of which are affected by the choice of the upper boundary condition (UBC), the altitude of the exobase also depends on the choice of the UBC.

\begin{figure}[h!]
    \centering
    \includegraphics[scale=0.33]{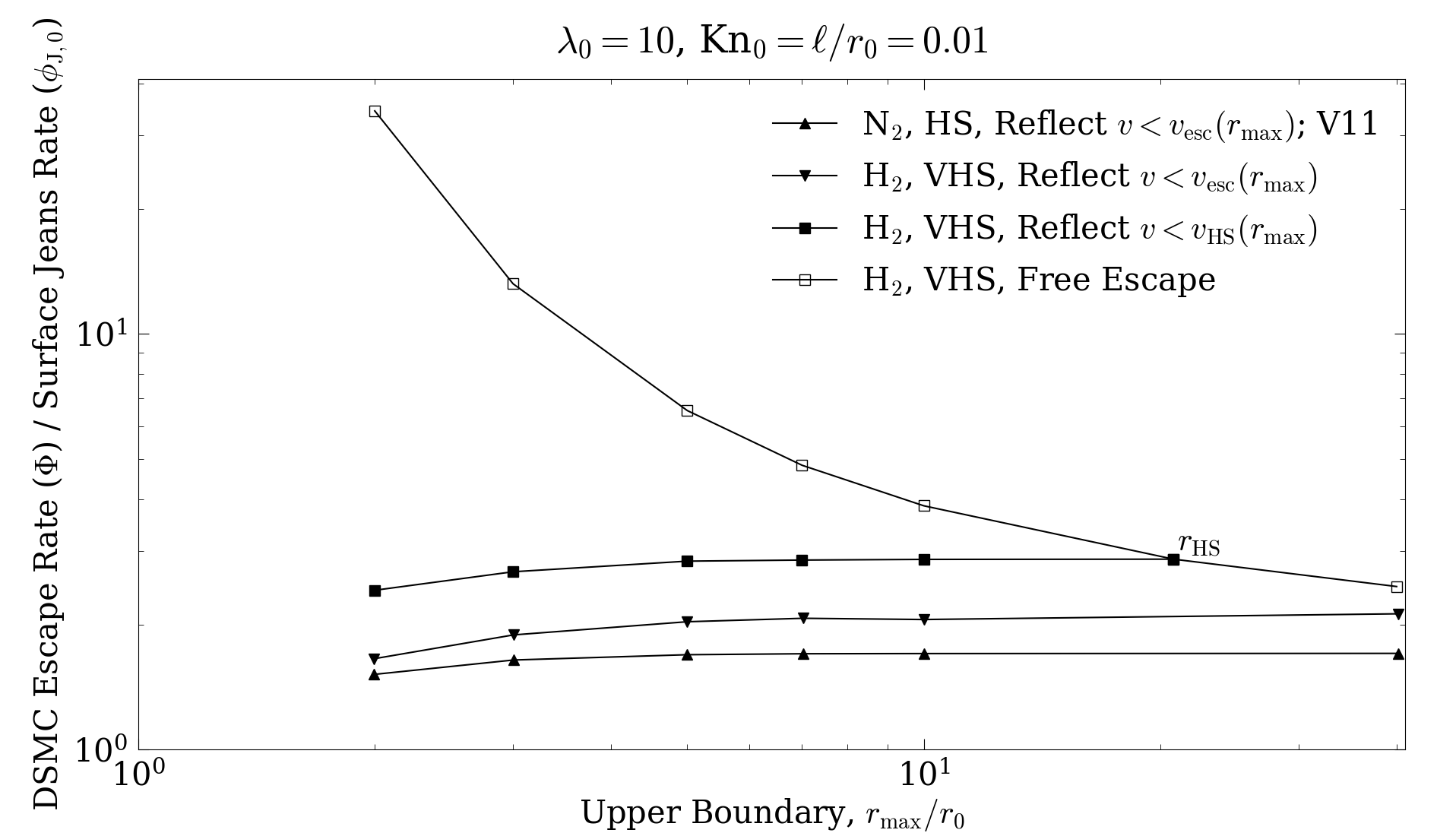}
    \caption{DSMC escape rates, $\Phi$, versus the placement of the upper boundary ($r_\mathrm{max}$) normalized to the Jeans rate, $\phi_{\mathrm{J},0}$ at the lower boundary, $r_0$: $\phi_\mathrm{J, 0} = \pi r_0^2 n_0 v_0 \exp (- \lambda_\mathrm{J,0}) (1+\lambda_\mathrm{J,0})$. Triangles are from V11 for N$_2$ molecules and hard sphere (HS) collisions using UBC (1). The inverted triangles, squares, and diamonds are from this study for UBC's (1),(2), and (4), respectively, for H$_2$ molecules at Callisto and the variable hard sphere (VHS) collision model. The escape rates presented from this study represent steady-state values, where steady-state was reached in $\sim$10$^5$ time-steps, $\sim$3$\times 10^5$ s.}
    \label{fig:volkov_comp}
\end{figure}

Fig. \ref{fig:volkov_comp} displays the DSMC escape rate, $\Phi$, normalized to the surface Jeans escape rate, $\phi_\mathrm{J, 0} = \pi r_0^2 n_0 \sqrt{ \frac{8 k T_0}{\pi m}} \exp (- \lambda_\mathrm{J,0}) (1+\lambda_\mathrm{J,0})$ vs. $r_{max}$ for UBC's (1), (2), and (4), where $\lambda_\mathrm{J, 0} = 10$ and Kn$_0(r_0) = \ell(r_0) / r_0 = 0.01$. These rates are compared to a simulation result in V11 (Fig. 4c therein), which illustrates the influence of the placement of $r_\mathrm{max}$ on $\Phi / \phi_\mathrm{J, 0}$. As can be seen, the placement of the upper boundary can affect the escape rates for all of the cases as collisions in the exosphere can induce escape (e.g., V11). With the exception of UBC (4), as the upper boundary increases, the physical space in which exospheric collisions occur increases, resulting in larger escape rates than when the domain is truncated at lower altitudes. The opposite is, of course, true for UBC (4), as the escape rate converges to UBC (2) when the upper boundary is at Callisto's Hill sphere where all particles are lost or for a much larger Hill sphere, the difference becomes only $\sim$1.2 that of UBC (1) by $r_\mathrm{max} / r_0 = 40$. As expected in these simulations, escape to the Hill sphere, UBC (2), is larger than escape when implementing UBC (1) due to the diminished threshold of $v_\mathrm{HS}$ relative to $v_\mathrm{esc}$. To emphasize the usefulness of scaling, we also compare our results to a result in V11 for an N$_2$ atmosphere with the same $\lambda_\mathrm{J, 0}$ and Kn$_0(r_0)$ using UBC (1). The trend with increasing $r_{max}$ is close to that for H$_2$ at Callisto when also using UBC (1). Within numerical uncertainties, the difference in magnitude, $\sim 10-20 \%$, is due to the use of the HS collision scheme in V11, rather than the collision scheme used here (VHS and LB) which approximates the dependence on the particles' relative speed and internal degrees of freedom; see \cite{tucker2012} for more details.

\begin{figure}[t]
    \centering
    \includegraphics[scale=0.28]{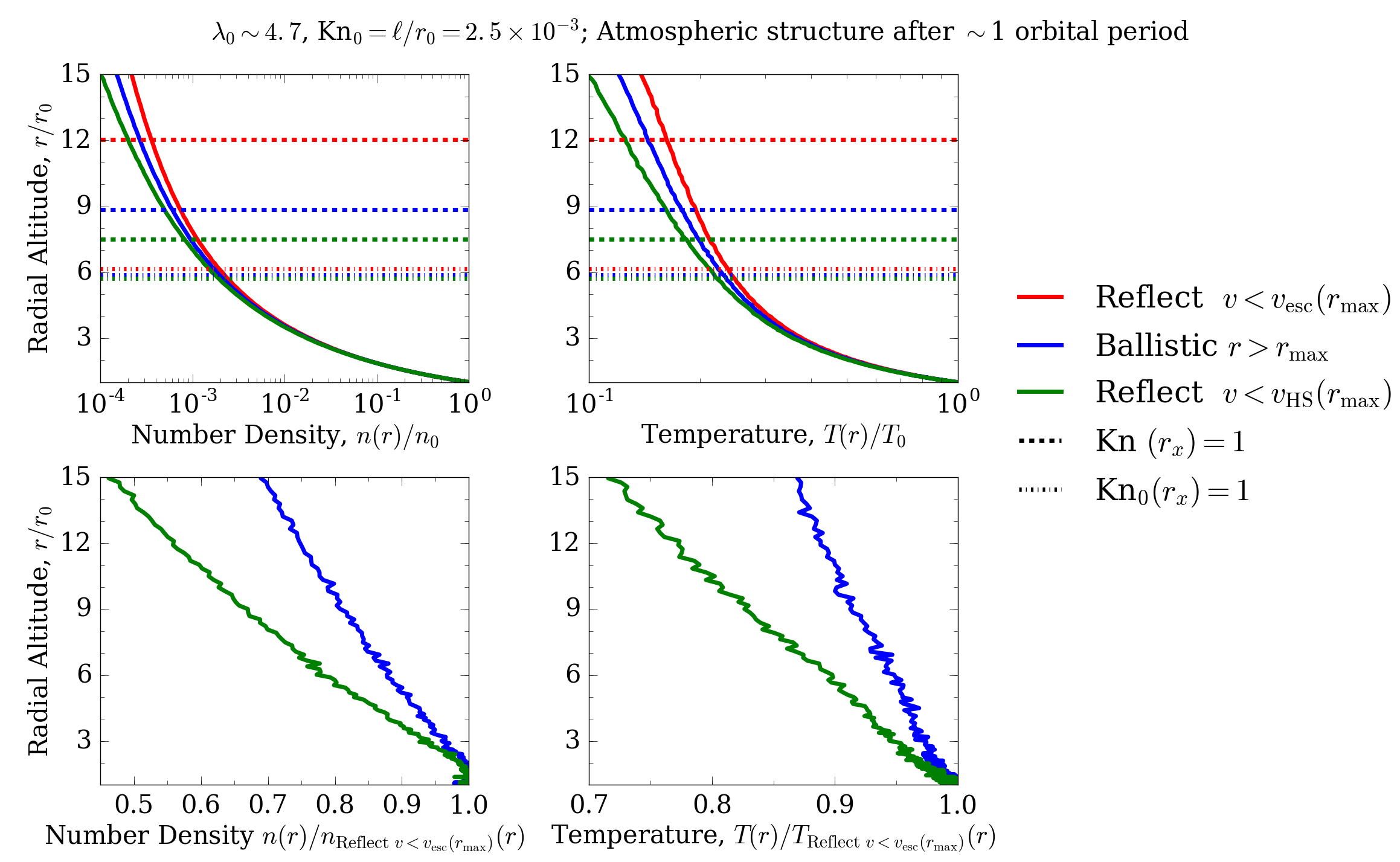}
    \caption{Top panels: Radial profiles of (\textit{left}) number density, $n(r)$, and (\textit{right}) temperature, $T(r)$, normalized by their surface values, $n_0$ and $T_0$ (see Table \ref{tab:exobase_params}), vs. radial altitude, $r/r_0$, for UBC (1, red solid lines), (2, green solid lines), and (3, blue solid lines). The exobase radii, calculated as Kn$(r_x)$ $=\ell(r_x) / H(r_x) \sim 1$ (dashed lines) and Kn$_0(r_x)=\ell(r_x) / r_0 \sim 1$ (dash-dotted lines), are also plotted in the same colors for each case. Bottom panels: Radial profiles of (\textit{left}) $n(r)$ and (\textit{right}) $T(r)$ for the UBC (2, green) and (3, blue) normalized to those of UBC (1).}
    \label{fig:atm_struct}
\end{figure}

\begin{figure}[h!]
    \centering
    \includegraphics[scale=0.28]{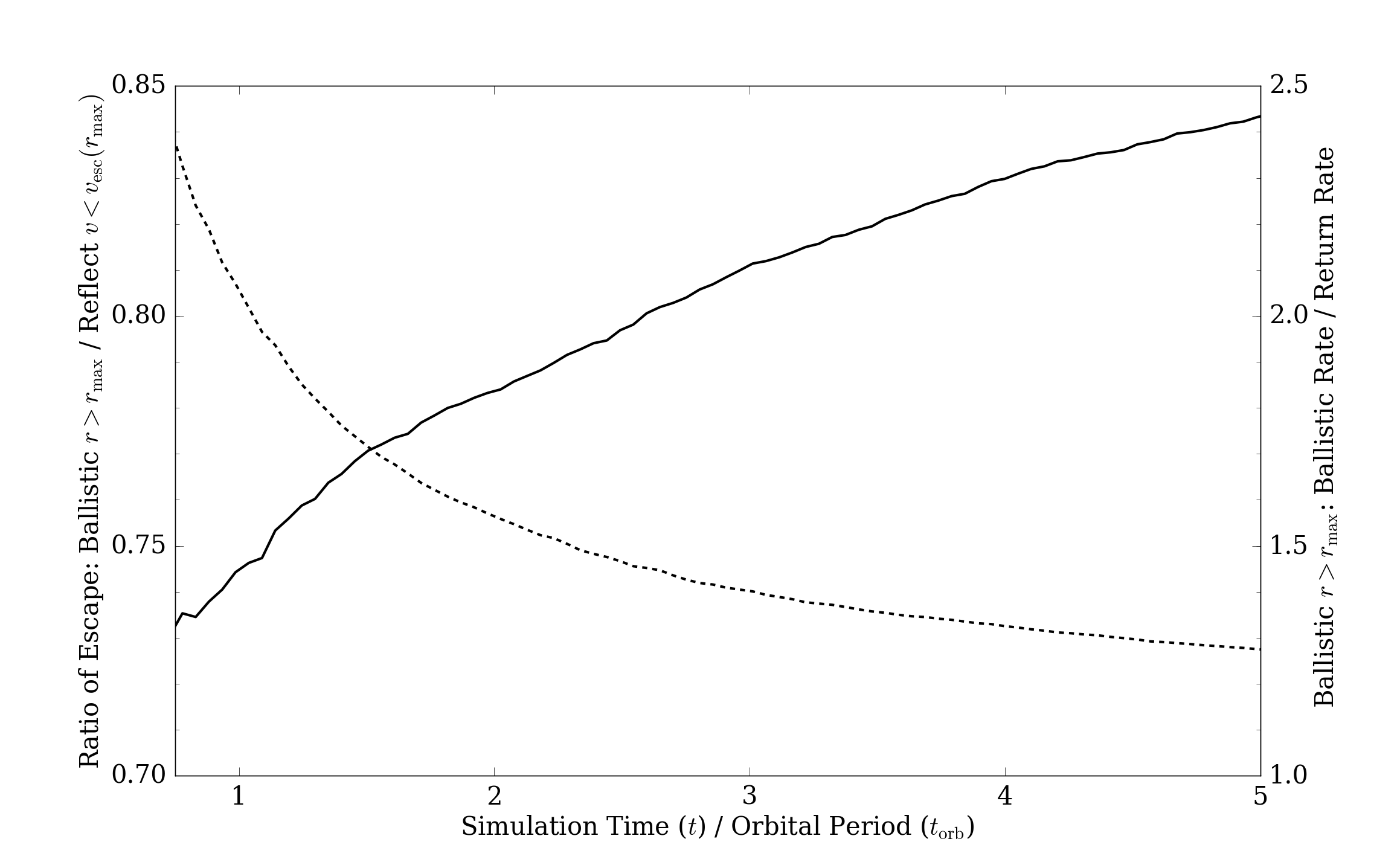}
    \caption{The ratio between the escape rate in the Ballistic $r > r_\mathrm{max}$, UBC (3), and Reflect $v < v_\mathrm{esc}(r_\mathrm{max})$, UBC (1) (solid line; y-axis: \textit{left}) and the ratio between the rate at which particles with insufficient escape trajectories reach $r_\mathrm{max}$ and subsequently follow ballistic trajectories (i.e., the ``Ballistic Rate'') and the rate at which those particles return to the domain (i.e., the ``Return Rate'') for the Ballistic $r > r_\mathrm{max}$ case (dashed line; y-axis: \textit{right}) vs. simulation time, $t$, normalized by Callisto's orbital period, $t_\mathrm{orb} \sim 1.44 \times 10^6$ s.}
    \label{fig:ball_norm}
\end{figure}

Fig. \ref{fig:atm_struct} shows how UBC's (1), (2) and (3) affect the structure of the upper atmosphere, with source parameters and results summarized in Table \ref{tab:exobase_params}. Starting from vacuum, steady-state is attained in the atmospheres where UBC's (1) and (2) were implemented within the first few averages taken, $\sim$10$^5$ steps ($\sim$7.5$\times 10^4$ s); whereas the structure of the atmosphere in which UBC (3) was implemented continued to change even after 1 $t_\mathrm{orb}$. The top left and right panels of Fig. \ref{fig:atm_struct} compare the simulated density and temperature profiles, respectively, with the exobase radii, calculated as Kn$(r_x) \sim 1$ and Kn$_0(r_x) \sim 1$, also indicated for each case. Since the magnitude differences in the profiles in the upper panels are difficult to distinguish on a logarithmic scale, profiles for UBC's (2) and (3) are scaled to those for UBC (1) in the bottom panels for clarity. Since escape is larger for UBC (2) than for UBC (1), the exosphere in the former is less dense and cooler. However, a surprising result is the difference in the profile for UBC (1) from that for UBC (3), in which particles' ballistic motion above $r_\mathrm{max}$ is fully tracked. This is further elaborated on in Fig. \ref{fig:ball_norm}, which compares the atmospheric escape rates as well as the ratio between the rates at which particles begin to follow ballistic trajectories above $r_\mathrm{max}$ and at which those particles return to the domain for these cases up to 5 $t_\mathrm{orb}$. Even by this time in the simulation a non-negligible fraction of the ballistic particles above $r_\mathrm{max}$ have not yet returned to the domain, contrary to what is assumed when specular reflection is implemented. Therefore, although the escape rate $\Phi$ for UBC (3) is less than that for UBC (1) at $\sim$1 $t_\mathrm{orb}$, these ballistic particles would effectively contribute to the \textit{total} loss resulting in a less dense and cooler exosphere.

\setlength{\tabcolsep}{10pt}
\renewcommand{\arraystretch}{1.5}
\begin{table}[h!]
\centering
\caption{Macroscopic Parameters at 1 Orbital Period, $t_\mathrm{orb} \sim 1.4 \times 10^6$ s}
\begin{tabular}{|l|c|c|c|c|c|c|}
\hline
\multirow{3}{*}{Normalized Exobase Results} & \multicolumn{2}{c|}{UBC (1)$^a$:\tnote{a}} & \multicolumn{2}{c|}{UBC (2)$^a$:} & \multicolumn{2}{c|}{UBC (3)$^a$:} \\
& \multicolumn{2}{c|}{Reflect} & \multicolumn{2}{c|}{Reflect} & \multicolumn{2}{c|}{Ballistic}\\
& \multicolumn{2}{c|}{$v < v_\mathrm{esc}(r_\mathrm{max})$} & \multicolumn{2}{c|}{$v< v_\mathrm{HS}(r_\mathrm{max})$} & \multicolumn{2}{c|}{$r > r_\mathrm{max}$}\\
\hline
Radius$^b$, $r_x/r_0$ \tnote{b} & 12.0 & 6.14 & 7.51 & 5.70 & 8.85 & 5.88 \\
\hline
Number density$^b$, $n_x/n_0$ $\left( 10^{-4} \right)$ & 3.58 & 19.5 & 8.18 & 19.4 & 6.06 & 19.4 \\
\hline
Temperature$^b$, $T_x/T_0$ $\left( 10^{-1} \right)$ & 1.62 & 2.38 & 1.83 & 2.22 & 1.78 & 2.31\\
\hline
Jeans parameter$^b$, $\lambda_{\mathrm{J}, x} / \lambda_\mathrm{J,0}$ $\left( 10^{-1} \right)$ & 5.12 & 6.83 & 7.27 & 7.88 & 6.35 & 7.36 \\
\hline
Jeans escape rate$^b$, $\phi_{\mathrm{J,} x} / \phi_\mathrm{J, 0}$ $\left( 10^{-2} \right)$ & 12.3 & 11.8 & 5.50 & 6.64 & 7.75 & 8.75 \\
\hline
DSMC escape rate$^c$, $\Phi / \phi_{\mathrm{J,} x}$ \tnote{c} & 2.36 & 2.47 & 6.16 & 5.16 & 2.97 & 2.63\\
\hline
\end{tabular}
\begin{tablenotes}[flushleft]\footnotesize
\item[a] $^a$Values in the left and right columns are those calculated using Kn$(r_x)\sim 1$ and Kn$_0 (r_x)\sim 1$, respectively.
\item[b] $^b$Values calculated at the exobase ($_x$) are normalized to the source parameters ($_0$), which are from the H$_2$ model atmospheres of \cite{carberrymogan2020}: $r_0 = 2410$ km, $n_0 = 4 \times 10^8$ cm$^{-3}$, $T_0 = 155$ K, $\lambda_\mathrm{J, 0} = \frac{G M_C m}{k T_0 r_0} \sim 4.7$, $\phi_\mathrm{J, 0} = \pi r_0^2 n_0 \sqrt{\frac{8 k T_0}{\pi m}} (1 + \lambda_\mathrm{J,0}) \exp (-\lambda_\mathrm{J,0}) \sim 4.9 \times 10^{29}$ s$^{-1}$. Here $M_C= 1.08 \times 10^{23}$ kg is the mass of Callisto and $m = 2.02$ amu is the mass of a H$_2$ molecule.
\item[c] $^c$DSMC escape rates, $\Phi$, for UBC's (1), (2), and (3) at 1 $t_\mathrm{orb}$ are $\sim$1.4$\times 10^{29}$ s$^{-1}$, $\sim$1.7$\times 10^{29}$ s$^{-1}$, and $\sim$1.1$\times 10^{29}$ s$^{-1}$ respectively.
\end{tablenotes}
\label{tab:exobase_params}
\end{table}

The conditions at the upper boundary and the resulting upper atmospheric structures also affect the location of the nominal exobase. As can be seen, when Kn$_0 (r_x) \sim 1$ (dash-dotted lines), the exobase altitudes for all three cases are similar (see values in Table \ref{tab:exobase_params}). From the bottom left panel of Fig. \ref{fig:atm_struct} the densities at $r_x \sim 6 r_C$ for UBC's (2) and (3) differ by $\sim$10 -- 20$\%$ of that for UBC (1), so the differences in densities at the exobase are not huge. However, this is not the case for Kn$(r_x)\sim 1$ (dashed lines), typically used in Jeans escape approximation due to the differences in the local temperature. As seen in the top panels of Fig. \ref{fig:atm_struct}, the nominal exobase can differ by several radii depending on the UBC implemented.

For the surface parameters in Fig. \ref{fig:atm_struct}, Table \ref{tab:exobase_params} lists the relation between $\lambda_\mathrm{J}$ and $\phi_\mathrm{J}$, among other parameters, calculated at Kn$(r_x) = 1$ and Kn$_0 (r_x) = 1$ revealing that, although $r_x$ and the properties calculated at $r_x$ differ significantly, the Jeans escape rates are not so different. However, the calculated DSMC rates, $\Phi$, can differ significantly from $\phi_\mathrm{J, x}$ depending on the UBC. Indeed for UBC's (1), (2), and (3) $\phi_{\mathrm{J}, x}$ is $\sim$40$\%$, $\sim$20$\%$ and $\sim$35$\%$ that of $\Phi$. Similarly, despite the large difference in $r_x$ (e.g., for UBC (1) Kn$(r_x) \sim 1$ occurs at $\sim$12$r_0$, whereas Kn$_0 (r_x) \sim 1$ occurs at $\sim$6$r_0$) the location where collisions become so infrequent that they can no longer maintain thermal equilibrium and, as a result, individual temperature components begin to diverge from one another occur around 2 scale heights below $r_x$. That is, although $r_x$ significantly differ, so too does the corresponding scale height ($H(r_x) \sim 5 r_0$ for Kn$(r_x) = 1$, whereas $H(r_x) \sim 2 r_0$ for Kn$_0(r_x) = 1$), leading to a similar radial location at which equilibrium begins to break down well below $r_x$ (e.g., \cite{carberrymogan2020, tucker2013}). This result is true for UBC's (1), (2), and (3), although the nature of the individual temperature components can differ. For example, due to the diminished escape threshold and hence enhanced escape rate in UBC (2) relative to UBC (1), whereas the average translational and transverse temperatures of the former become cooler than those of the latter with increasing altitude, the average radial temperature of the latter becomes cooler than that of the former.

Finally, as mentioned earlier, after $\sim t_\mathrm{orb}$, a significant fraction of the particles following ballistic trajectories above $r_\mathrm{max}$ still have not returned to the domain, affecting the structure of the upper atmosphere seen in Fig. \ref{fig:atm_struct}. Fig. \ref{fig:ball_norm} indicates the effect on the escape rate due to the time it takes for ballistic particles to return to the domain as compared to immediate specular reflection at $r_{max}$, UBC (1). The returning particles slowly refill the exosphere, where they induce escape via collisions, enhancing the calculated escape rate, which gradually approaches the rate for UBC (1). Fig. \ref{fig:ball_norm} also shows the ratio between the rate at which particles reach $r_\mathrm{max}$ and follow ballistic trajectories thereafter (the ``Ballistic Rate'') and the rate at which those ballistic particles return to the domain (the ``Return Rate''). When specular reflection is implemented, this ratio is assumed to be unity; that is, all particles that follow ballistic trajectories above $r_\mathrm{max}$ return to the domain. However, even after 5$t_\mathrm{orb}$, this is still not the case when the ballistic trajectories are explicitly tracked and, thus, the Ballistic Rate is still greater the Return Rate. If the simulation were run for much longer times, in the absence of external processes, the escape rate for UBC (3) relative to UBC (1) and the Ballistic and Return Rates are expected to converge. However, such results are not very useful when other dynamical timescales are of greater importance; e.g., at 5$t_\mathrm{orb}$ the escape rate for UBC (3) is still $<85\%$ of that for UBC (1).

\section{Summary}

1D DSMC calculations of even multi-component atmospheres modified by external processes are now at the point where very useful results can be obtained by simulations on a modern desktop computer. Although more extensive, multi-dimensional simulations can be carried out by parallel processing, the simple 1D simulations can be extremely useful in addressing issues and physical processes discussed in the extensive literature from the work of Jeans \citep{jeans1921dynamical} until the present on accurately treating a planet's transition region. This is the region that evolves from highly collisional to collisionless and determines the effect on the planet's atmospheric evolution due to escape. Because of the continuing usefulness of such simulations and their present efficiency, we reviewed issues affected by the choice of the upper boundary conditions (UBC) that are often implemented in order to limit run times. In addition we briefly reminded the reader of both the usefulness of and limits on scaling using the value of the Jeans parameter and Knudsen number at the lower boundary as a means of expanding the usefulness of simulation results as well as the effect on those results of the model for implementing collisions. These issues are more extensively discussed elsewhere as indicated above. We did not address the issues at the lower boundary. By implementing an upward flux boundary condition, the lower boundary could be in a collisional region at some altitude at which the atmospheric properties are known or, as for the thin atmospheres on the outer solar system bodies, at their physical surface. Whereas the lower boundary for the former is readily implemented, the role of the surface for thin atmospheres is critical and differs significantly between bodies.

Here we described how the choice of the UBC in such simulations affects the results of an atmospheric simulation. Because the collision rate decreases with increasing altitude (radial distance from the surface), depending on the goal of the simulations, at some point collisions can be ignored. But since objects of interest rotate and/or orbit large bodies and have atmospheres comprised of a variety of species, the time scale required to reach steady-state is critical. Here we focused on the dynamical time scale using a relatively long-lived species, H$_2$, at Callisto as an example for which a truly steady-state exosphere never occurs in an orbital period. That is, although specular reflection can be accurate over long dynamical time scales, for the example considered here, it overestimates the escape rate by $\sim$25$\%$ in times less than 1 Callisto orbit. Although the effect of the UBC is much less critical for molecules with relatively short lifetimes, for those with long lifetimes and/or capable of escaping the atmosphere, the choice of the UBC should be determined by the dynamical time scale as well as any external and chemical kinetics effects (e.g., photodissociation lifetimes) that can be readily implemented. Therefore the results presented are intended to be a guide when implementing much more detailed 1D DSMC simulations.  

\section*{Acknowledgments} This work is supported by grant 80NSSC20M0193 from NASA Goddard Space Flight Center’s Solar System Exploration Division. This research was carried out on the High Performance Computing resources at New York University Abu Dhabi.

\bibliography{CarberryMoganEtAl_2021.bib}

\begin{thebibliography}{}

\bibitem[Austin and Goldstein, 2000]{austin2000}
Austin, J.~V. and Goldstein, D.~B. (2000).
\newblock Rarefied gas model of {I}o's sublimation-driven atmosphere.
\newblock {\em Icarus}, 148(2):370--383.

\bibitem[Bird, 1994]{bird1994}
Bird, G.~A. (1994).
\newblock {\em Molecular gas dynamics and the direct simulation of gas flows}.
\newblock Clarendon press Oxford.

\bibitem[Brinkmann, 1970]{brinkmann1970}
Brinkmann, R.~T. (1970).
\newblock Departures from {J}eans' escape rate for {H} and {H}e in the
  {E}arth's atmosphere.
\newblock {\em Planet. and Space Sci.}, 18(4):449--478.

\bibitem[Carberry~Mogan et~al., 2020]{carberrymogan2020}
Carberry~Mogan, S.~R., Tucker, O.~J., Johnson, R.~E., Sreenivasan, K.~R., and
  Kumar, S. (2020).
\newblock The influence of collisions and thermal escape in {C}allisto's
  atmosphere.
\newblock {\em Icarus}, 352:113932.

\bibitem[Carberry~Mogan et~al., 2021]{carberrymogan2021}
Carberry~Mogan, S.~R., Tucker, O.~J., Johnson, R.~E., Vorburger, A., Galli, A.,
  Marchand, B., Tafuni, A., Sahin, I., Sreenivasan, K.~R., and Kumar, S.
  (2021).
\newblock A tenuous, collisional atmosphere on {C}allisto.
\newblock {\em Icarus}.

\bibitem[Combi, 1996]{combi1996}
Combi, M.~R. (1996).
\newblock Time-dependent gas kinetics in tenuous planetary atmospheres: The
  cometary coma.
\newblock {\em Icarus}, 123(1):207--226.

\bibitem[Erwin et~al., 2013]{erwin2013}
Erwin, J., Tucker, O.~J., and Johnson, R.~E. (2013).
\newblock Hybrid fluid/kinetic modeling of {P}luto's escaping atmosphere.
\newblock {\em Icarus}, 226(1):375--384.

\bibitem[Hill, 1878]{hill1878}
Hill, G.~W. (1878).
\newblock Researches in the lunar theory.
\newblock {\em Am. J. of Math.}, 1(1):5--26.

\bibitem[Jeans, 1921]{jeans1921dynamical}
Jeans, J. (1921).
\newblock {\em The dynamical theory of gases}.
\newblock University Press.

\bibitem[Johnson et~al., 2015]{johnson2015}
Johnson, R.~E., Oza, A., Young, L.~A., Volkov, A.~N., and Schmidt, C. (2015).
\newblock Volatile loss and classification of {K}uiper belt objects.
\newblock {\em Astrophys. J.}, 809(1):43.

\bibitem[Johnson et~al., 2013]{johnson2013}
Johnson, R.~E., Volkov, A.~N., and Erwin, J.~T. (2013).
\newblock Molecular-kinetic simulations of escape from the ex-planet and
  exoplanets: Criterion for transonic flow.
\newblock {\em Astrophys. J. Letters}, 768(1):L4.

\bibitem[Larsen and Borgnakke, 1974]{larsen1974}
Larsen, P.~S. and Borgnakke, C. (1974).
\newblock Statistical collision model for simulating polyatomic gas with
  restricted energy exchange.
\newblock In {\em Rarefied Gas Dynamics, 9th Symposium}.

\bibitem[Leblanc et~al., 2017]{leblanc2017}
Leblanc, F., Oza, A.~V., Leclercq, L., Schmidt, C., Cassidy, T., Modolo, R.,
  Chaufray, J.~Y., and Johnson, R.~E. (2017).
\newblock On the orbital variability of {G}anymede's atmosphere.
\newblock {\em Icarus}, 293:185--198.

\bibitem[Leclercq et~al., 2020]{leclercq2020}
Leclercq, L., Williamson, H.~N., Johnson, R.~E., Tucker, O.~J., Tian, L., and
  Snowden, D. (2020).
\newblock Molecular kinetic simulations of transient perturbations in a
  planet's upper atmosphere.
\newblock {\em Icarus}, 335:113394.

\bibitem[Lee et~al., 2015]{lee2015}
Lee, Y. et~al. (2015).
\newblock A comparison of 3-{D} model predictions of {M}ars' oxygen corona with
  early {MAVEN IUVS} observations.
\newblock {\em Geophys. Res. Letters}, 42(21):9015--9022.

\bibitem[Marconi, 2007]{marconi2007}
Marconi, M. (2007).
\newblock A kinetic model of {G}anymede's atmosphere.
\newblock {\em Icarus}, 190(1):155--174.

\bibitem[Merryfield and Shizgal, 1994]{merryfield1994}
Merryfield, W.~J. and Shizgal, B.~D. (1994).
\newblock Discrete velocity model for an escaping single-component atmosphere.
\newblock {\em Planet. and Space Sci.}, 42(5):409--419.

\bibitem[Moore et~al., 2009]{moore2009}
Moore, C.~H., Goldstein, D.~B., Varghese, P.~L., Trafton, L.~M., and Stewart,
  B. (2009).
\newblock 1-{D} {DSMC} simulation of {I}o's atmospheric collapse and
  reformation during and after eclipse.
\newblock {\em Icarus}, 201(2):585--597.

\bibitem[Shematovich et~al., 2005]{shematovich2005}
Shematovich, V.~I., Johnson, R.~E., Cooper, J.~F., and Wong, M.~C. (2005).
\newblock Surface-bounded atmosphere of {E}uropa.
\newblock {\em Icarus}, 173(2):480--498.

\bibitem[Tucker et~al., 2012]{tucker2012}
Tucker, O.~J., Erwin, J.~T., Deighan, J.~I., Volkov, A.~N., and Johnson, R.~E.
  (2012).
\newblock Thermally driven escape from {P}luto's atmosphere: A combined
  fluid/kinetic model.
\newblock {\em Icarus}, 217(1):408--415.

\bibitem[Tucker et~al., 2013]{tucker2013}
Tucker, O.~J., Johnson, R.~E., Deighan, J.~I., and Volkov, A.~N. (2013).
\newblock Diffusion and thermal escape of {H}$_2$ from {T}itan's atmosphere:
  Monte {C}arlo simulations.
\newblock {\em Icarus}, 222(1):149--158.

\bibitem[Tucker et~al., 2021]{tucker2020}
Tucker, O.~J., Killen, R.~M., Johnson, R.~E., and Saxena, P. (2021).
\newblock Lifetime of a transient atmosphere produced by {L}unar volcanism.
\newblock {\em Icarus}, 359:114304.

\bibitem[Tucker et~al., 2016]{tucker2016}
Tucker, O.~J., Waalkes, W., Tenishev, V.~M., Johnson, R.~E., Bieler, A., Combi,
  M.~R., and Nagy, A.~F. (2016).
\newblock Examining the exobase approximation: {DSMC} models of {T}itan's upper
  atmosphere.
\newblock {\em Icarus}, 272:290--300.

\bibitem[Volkov et~al., 2011]{volkov2011a}
Volkov, A.~N., Tucker, O.~J., Erwin, J.~T., and Johnson, R.~E. (2011).
\newblock Kinetic simulations of thermal escape from a single component
  atmosphere.
\newblock {\em Phys. of Fluids}, 23(6):066601.

\bibitem[Walker et~al., 2010]{walker2010}
Walker, A.~C., Gratiy, S.~L., Goldstein, D.~B., Moore, C.~H., Varghese, P.~L.,
  Trafton, L.~M., Levin, D.~A., and Stewart, B. (2010).
\newblock A comprehensive numerical simulation of {I}o's sublimation-driven
  atmosphere.
\newblock {\em Icarus}, 207(1):409--432.

\bibitem[Walker et~al., 2012]{walker2012}
Walker, A.~C., Moore, C.~H., Goldstein, D.~B., Varghese, P.~L., and Trafton,
  L.~M. (2012).
\newblock A parametric study of {I}o's thermophysical surface properties and
  subsequent numerical atmospheric simulations based on the best fit
  parameters.
\newblock {\em Icarus}, 220(1):225--253.

\end{thebibliography}

\end{document}